\documentclass[12pt]{article}
\usepackage{authblk}
\usepackage{amsmath}
\usepackage{amssymb}
\usepackage{booktabs}
\usepackage{graphicx}
\usepackage{psfrag,epsf}
\usepackage{enumerate}
\usepackage{natbib}
\usepackage{url} % not crucial - just used below for the URL
\usepackage{comment}
\RequirePackage[colorlinks,citecolor=blue,urlcolor=blue]{hyperref}
\usepackage{setspace}

\allowdisplaybreaks

\usepackage[pagewise]{lineno}
\newcommand*\patchAmsMathEnvironmentForLineno[1]{%
        \expandafter\let\csname old#1\expandafter\endcsname\csname 
        #1\endcsname
        \expandafter\let\csname oldend#1\expandafter\endcsname\csname 
        end#1\endcsname
        \renewenvironment{#1}%
        {\linenomath\csname old#1\endcsname}%
        {\csname oldend#1\endcsname\endlinenomath}}%
\newcommand*\patchBothAmsMathEnvironmentsForLineno[1]{%
        \patchAmsMathEnvironmentForLineno{#1}%
        \patchAmsMathEnvironmentForLineno{#1*}}%
\AtBeginDocument{%
        \patchBothAmsMathEnvironmentsForLineno{equation}%
        \patchBothAmsMathEnvironmentsForLineno{align}%
        \patchBothAmsMathEnvironmentsForLineno{flalign}%
        \patchBothAmsMathEnvironmentsForLineno{alignat}%
        \patchBothAmsMathEnvironmentsForLineno{gather}%
        \patchBothAmsMathEnvironmentsForLineno{multline}%
}

\newcommand{\blind}{0}

\begin{document}

\def\spacingset#1{\renewcommand{\baselinestretch}%
{#1}\small\normalsize} \spacingset{1}

%%%%%%%%%%%%%%%%%%%%%%%%%%%%%%%%%%%%%%%%%%%%%%%%%%%%%%%%%%%%%%%%%%%%%%%%%%%%%%

\title{\bf Nonparametric Block Bootstrap Kolmogorov-Smirnov Goodness-of-Fit Test}
\if0\blind
{
  \author[1,2]{Mathew Chandy}
  \author[1]{Elizabeth D. Schifano}
  \author[1]{Jun Yan\thanks{
    The authors gratefully acknowledge the National Science Foundation (NSF).}\hspace{.2cm}}
  \author[3]{Xianyang Zhang$^{\ast}$ }
  \affil[1]{Department of Statistics, University of Connecticut}
  \affil[2]{Department of Statistics and Data Science, University of California, Los Angeles}
  \affil[3]{Department of Statistics, Texas A\&M University}
  \maketitle
} \fi

\if1\blind
{
  \bigskip
  \bigskip
  \bigskip
  \author{Anonymous Author}
} \fi

\maketitle

\bigskip
\begin{abstract}
The Kolmogorov--Smirnov (KS) test is a widely used statistical test that
assesses the conformity of a sample to a specified distribution. Its efficacy,
however, diminishes with serially dependent data and when parameters
within the hypothesized distribution are unknown. For independent data,
parametric and nonparametric bootstrap procedures are available to adjust for
estimated parameters. For serially dependent stationary data, parametric
bootstrap has been developed with a working serial dependence structure. A
counterpart for the nonparametric bootstrap approach, which needs a bias
correction, has not been studied. Addressing this gap, our study introduces a
bias correction method employing a nonparametric block bootstrap, which
approximates the distribution of the KS statistic in assessing the
goodness-of-fit of the marginal distribution of a stationary series,
accounting for unspecified
serial dependence and unspecified parameters. We assess its effectiveness
through simulations, scrutinizing both its size and power. The practicality of
our method is further illustrated with an examination of stock returns from the
S\&P 500 index, showcasing its utility in real-world
applications.
\end{abstract}

\noindent%
{\it Keywords:}  bias-correction, financial data, marginal distribution,
  stationary series, time series
\vfill

\newpage

\doublespacing

\section{Introduction}\label{sec:intro}

The standard one-sample Kolmogorov--Smirnov (KS) test is widely
recognized as an effective good-of-fit test for continuous distributions.
Consider $X_1,  \cdots , X_n$, a random sample of size~$n$ from some continuous
distribution and the null hypothesis $H_0$ that $X_i$'s follow a specific
hypothesized distribution~$F(x; \theta_0)$, where $\theta_0$ are the
specified parameters
of the hypothesized distribution family.
If we let $F_n(t) = \sum_{i=1}^n \mathbf{1}(\{X_i \le t\} / n$ be the empirical cumulative
distribution function of the sample, where $\mathbf{1}\{\cdot\}$ is the indicator
function, the KS test statistic takes the form
\[
  D_n = \sqrt{n} \sup_x | F_{n}(x) - F(x; \theta_0)|.
\]
As sample size $n\to \infty$, the distribution of $D_n$ converges to that of the
absolute value of standard Brownian bridge, which is known as the Kolmogorov
distribution \citep{stephens1974edf}. This distribution function is
precisely computable using contemporary statistical software
\citep{marsaglia2003evaluating}. The versatility of the KS test allows its
application across various domains, such as analyzing cosmic microwave
background radiation \citep{naess2012application}, monitoring the count rate of
radioactive data \citep{aslam2020introducing}, gear condition monitoring
\citep{andrade2001gear}, studying images of breast cancer tumors
\citep{demidenko2004kolmogorov}, and examining financial markets
\citep{lux2001turbulence}.

Despite its widespread use, the KS test can be
misapplied when its foundational assumptions are overlooked. The KS test
assumes that the data are independently and identically distributed and
that the hypothesized distribution is continuous and fully specified without
the need for parameter estimation. \citet{zeimbekakis2024misuses} examines
common misapplications of the one-sample KS test. One notable inappropriate
use is when the hypothesized distribution contains unspecified parameters.
In this case, the tests are generally constructed by substituting the unknown
parameters by their estimates, and the asymptotic null distribution of
the test statistic may depend in a complex way on the unknown parameters.
The problem of unspecified parameters can be handled by a parametric
bootstrap, where bootstrap samples of the test statistics are constructed from
samples generated from the fitted hypothesized distribution.
Alternatively, \citet{babu2004goodness} address this issue through a
nonparametric bootstrap (NPB) method that corrects the bias in the asymptotic null
distribution.

Another prevalent misapplication of the KS test arises when data exhibit serial
dependence, which is often overlooked in analyses. For example, in an
investigation of performance variation in high-performance computing systems,
\citet{tuncer2019ieee} did not describe how or if they accounted for serial
dependence when applying the two-sample KS test.
A particularly complex scenario arises when the distribution has unspecified
parameters, and the data are serially dependent.
The distribution of the test statistic under the null hypothesis depends on both
the unknown parameters and unknown serial dependence.
\citet{zeimbekakis2024misuses} recommend a semiparametric bootstrap (SPB)
approach, where the parameters of the hypothesized distribution and the serial
dependence both need to be estimated and their uncertainty accounted for.
Nonetheless, if the working dependence model is too far from the truth, the
results of the tests would not be reliable. For this reason, a nonparametric
method that does not specify the serial dependence is desired.

Developing a fully nonparametric solution for the KS
tests in the presence of serial dependence presents a significant
challenge. The null distribution's characteristics are intricately linked to
the structure of the serial dependence, which can vary widely in practical
situations. Employing a semiparametric method
\citep{zeimbekakis2024misuses} would necessitate defining a
specific model for this dependence, despite the primary focus being on
evaluating the marginal distribution of a stationary series. To date, a bias
correction method for block bootstrap, akin to the NPB
approach of \citet{babu2004goodness} for independent data,
has not been established. This study seeks
to bridge this gap by introducing a bias correction technique for the
nonparametric block bootstrap (NPBB), tailored for use in scenarios where the
hypothesized distribution has unspecified parameters and the data
exhibit serial dependence. Our approach reduces to that of
\citet{babu2004goodness} when the data are independent.

The remainder of this paper is structured as follows:
Section~\ref{sec:methods} provides an overview of the block bootstrap
procedure and introduces the bias correction methodology for the
NPBB KS test. Section~\ref{sec:simu} is divided
into two parts; initially, we evaluate the NPBB KS test's ability to maintain
its size, i.e., its consistency in not rejecting the null hypothesis when
it is indeed true. Subsequently, we examine the NPBB KS test's power, assessing
whether it can reject the null hypothesis when it is false. Practical
applications of our method are presented in Section~\ref{sec:real}, where
we apply the approach to analyze if S\&P 500 index stock return
data adheres to either the Normal or the Student's $t$ distribution. The
paper concludes with Section~\ref{sec:conclusion}, offering final thoughts
and remarks.

\section{Methods}\label{sec:methods}

Consider a stationary time series $\{X_i: i = 1, \ldots, n\}$ with length~$n$.
We are interested in testing whether or not $X_i$ follows a distribution in a
parametric family of distribution~$F$ indexed by a parameter
vector~$\theta$. That is, the null hypothesis is
\[
  H_0: X_i \sim F(\cdot ; \theta), \quad i = 1, \ldots, n,
\]
for some unspecified parameter $\theta$.
The alternative hypothesis $H_A$ is that the marginal distribution of $X_i$ does
not follow~$F$ for any parameter value~$\theta$. This is a challenging situation
because both the parameters and the serial dependence structure are unknown.

\subsection{Kolmogorov--Smirnov test}

First, let us review how the KS statistic is computed for an independent
sample with fitted parameters. Let $\hat\theta_n$ denote the parametrically
fitted parameters, which can be obtained from any consistent estimator with
asymptotic normality property.  Let $F_n$ be the empirical distribution
function based on $X_1, \cdots ,X_n$.
Define
\begin{equation*}
Y_n(x; \hat\theta_n) = \sqrt{n}(F_n(x) - F(x; \hat\theta_n)).
\end{equation*}
Then, the one-sample KS goodness of fit statistic is
\begin{equation*}
T_n := \sup_x|Y_n(x; \hat\theta_n)|.
\end{equation*}

Because of the estimation uncertainty in $\hat\theta_n$, the asymptotic
distribution of $Y_n$ is no longer distribution-free. This is in contrast to the
case where $\theta$ is known.  Note that
\begin{equation*}
Y_n(x; \hat\theta) = \sqrt{n}(F_n(x) - F(x; \theta_0)) -
\sqrt{n}(F(x; \hat\theta_n) - F(x; \theta_0)),
\end{equation*}
where $\theta_0$ is the true parameter value of $\theta$ under~$H_0$.
When $\hat\theta_n$ is replaced with $\theta_0$, as in the standard KS testing
situation, the second term vanishes, and $T_n$ converges in distribution to that
of $\sup_t | J(F(t; \theta_0)) |$, where $J(\cdot)$ is the standard Brownian bridge
\citep{kolmogorov1933sulla}. With $\hat\theta_n$ in place of $\theta$, the
second term leads to a bias in the Brownian bridge, which was addressed by
\citet{babu2004goodness}.

When $\{X_i: i = 1, \ldots, n\}$ are stationary with serial
dependence, the statistic
$T_n$ still provides a reasonable measure for deviation from the null
hypothesis. In the next subsection, we review the NPB approach of
\citet{babu2004goodness} for independent data, which serves as the
basis for our extension to dependent series.

\subsection{NPB for independent data}

Continue with the case where $X_i$'s are independent but the parameters
are unspecified. Denote by $F^{(b)}_n$ the empirical distribution of the $b$th bootstrap sample and let
$\hat\theta^{(b)}_n$ be the parameter estimate based on the $b$th bootstrap
sample.
Using the bootstrap (asymptotic) theory, we can approximate the distribution of
$\sqrt{n}(F_n(x) - F(x; \theta_0))$ and
$\sqrt{n}(F(x; \hat\theta_n) - F(x; \theta_0))$
by that of $\sqrt{n}(F^{(b)}_n(x) - F_n(x))$ and
$\sqrt{n}(F(x; \hat\theta^{(b)}_n) - F(x; \hat\theta_n))$, respectively.
Therefore, if we define
\begin{align*}
Y^{(b)}_n(x) &= \sqrt{n}(F^{(b)}_n(x) - F_n(x)) -
               \sqrt{n}(F(x; \hat\theta^{(b)}_n) - F(x; \hat\theta_n)) \\
             &= \sqrt{n}(F^{(b)}_n(x) - F(x; \hat\theta^{(b)}_n)) -
               \sqrt{n}(F_n(x) - F(x; \hat\theta_n)),
\end{align*}
then $T^{(b)}_n := \sup_x|Y^{(b)}_n(x)|$ is the bootstrap statistic that is
expected to approximate the distribution of $T_n$.
We note that the term
$\sqrt{n}(F_n(x) - F(x; \hat\theta_n))$ is exactly the bias term considered in
\citet{babu2004goodness}.

In summary, the procedure of the NPB KS test for an independent sample is
summarized as follows. Repeat the following steps for $b \in \{1,  \cdots , B\}$.
\begin{enumerate}
\item
  Generate $X^{(b)}_1, \cdots ,X^{(b)}_n$ by sampling $X_1, \ldots, X_n$
  with replacement.
\item
  Obtain parametrically
  fitted parameters
	$\hat\theta^{(b)}_n$ of $\theta$ from $X^{(b)}_1, \cdots ,X^{(b)}_n$.
\item
  Obtain the empirical distribution function $F^{(b)}_n$ of
  $X^{(b)}_1, \cdots ,X^{(b)}_n$.
\item
  Calculate bootstrap KS statistic
  \[
    T^{(b)}_n = \sup_x \Big\vert \sqrt{n}\left(F^{(b)}_n(x)
    - F(x; \hat\theta^{(b)}_n)\right) - C_n(x) \Big\vert.
  \]
  where
  \[
    C_{n}(x) = \sqrt{n}\bigl(F_n(x) - F(x; \hat\theta_n)\bigr)
  \]
  is the estimated bias term.
\end{enumerate}

\citet{babu2004goodness} proves that $T^{(b)}_n$ and $T_n$ have the same
limiting distribution for almost all samples $X_1, \ldots, X_n$.
The p-value of the basic bootstrap KS test can be approximated
as $p = \sum_{b=1}^B \mathbf{1}\{T^{(b)}_n > T_n\} / B$.

\subsection{NPBB for stationary series}
\label{sub:npbb}

We now consider the case where $X_i$'s are realizations from a time series and
$X^{(b)}_1, \cdots ,X^{(b)}_n$ are generated by block bootstrap for
$b \in \{1, \ldots, B\}$.
Circular block-bootstrap can be done with overlapping blocks that can
wrap around the end of the series \citep{romano1992circular}.
Define blocks (assuming block size $l > 1$) as:
\begin{equation*}
Z_j =
    \begin{cases}
        \{X_j, \ldots, X_{j + l - 1}\}, & j = 1, \dots, n - l + 1,\\
        \{X_j, \ldots, X_n, X_1, \ldots, X_{j-n+l-1}\}, & j = n - l
        + 2 ,\dots, n.
    \end{cases}
\end{equation*}
A common
function for block size that is often considered optimal is
$l = \lceil n^{1/3} \rceil$ \citep{buhlmann1999block},
although \citet{hall1995blocking} and \citet{politis2004automatic} note that
the selection of block size should ideally take into account the strength
and direction of serial dependence. 
Now we draw $k$ blocks from the $(n - l + 1)$ blocks
of $Z_j$'s with replacement and then align them in the order they were picked to
form a block bootstrap sample. If $n$ is not a multiple of~$l$, the last block
selected will be reduced in size so that the final size of the block bootstrap
sample is $n$.

As in the independent case,
we use $F^{(b)}_n$ and $\hat\theta^{(b)}_n$ for the empirical distribution and
the estimated parameters based on the $b$th bootstrap sample,
$b = 1, \ldots, B$.
We then use block bootstrap to approximate the asymptotic distribution of
the KS statistic under $H_0$. Let
$\theta_0^* = \mathbb{E}^{*}[\hat\theta^{(b)}_n]$, where
$\mathbb{E}^{*}$ denotes the expectation with respect to the
bootstrap distribution (i.e., the randomness due
to the resampling using block bootstrap) conditional on the observations
$X_1, \dots, X_n$. In particular, we
use the distribution of $\sqrt{n}(F^{(b)}_n(x) - \mathbb{E}^{*}[F^{(b)}_n(x)])$
as an approximation of the distribution of
$\sqrt{n}(F_n(x) - F(x; \theta_0))$, and the distribution of
$\sqrt{n}(F(x; \hat\theta^{(b)}_n) - F(x; \theta_0^*))$ as
an approximation of the distribution of
$\sqrt{n}(F(x; \hat\theta_n) - F(x; \theta_0))$.

Then, for each $b \in \{1, \dots, B\}$, we can define
\begin{align*}
  Y^{(b)}_n(x) &= \sqrt{n}(F^{(b)}_n(x) - \mathbb{E}^{*}[F^{(b)}_n(x)]) -
             \sqrt{n}(F(x; \hat\theta^{(b)}_n) - F(x; \theta_0^*) \\
           &= \sqrt{n}(F^{(b)}_n(x) - F(x; \hat\theta^{(b)}_n)) -
             \sqrt{n}(\mathbb{E}^{*}[F^{(b)}_n(x)] - F(x; \theta_0^*)),
\end{align*}
and $T^{(b)}_n = \sup_x|Y^{(b)}_n(x)|$. Each $T_n^{(b)}$,
$b =1, \ldots, B$, is considered a draw from a distribution that approximates
the distribution of $T_n$. Therefore, the p-value of the observed statistic
$T_n$ can be assessed by positioning it against the empirical distribution of
$T_n^{(b)}$, $b = 1, \ldots, B$.

In summary, the procedure of the NPBB test is
summarized as follows. Repeat the following steps for $b \in \{1,  \cdots , B\}$.
\begin{enumerate}
\item
  Generate $X^{(b)}_1, \cdots ,X^{(b)}_n$ by applying circular block bootstrap
  on the original sample as
  defined previously.
\item
  Obtain parametrically fitted parameters
  $\hat\theta^{(b)}_n$ of $\theta$ from $X^{(b)}_1, \cdots ,X^{(b)}_n$.
\item
  Obtain the empirical distribution function $F^{(b)}_n$ of
  $X^{(b)}_1, \cdots ,X^{(b)}_n$.
\item
  Calculate bootstrap KS statistic
  \[
    T^{(b)}_n = \sup_x \Big\vert \sqrt{n}\left(F^{(b)}_n(x)
    - F(x; \hat\theta^{(b)}_n)\right) - K_n(x) \Big\vert.
  \]
  where
  \[
    K_{n}(x) = \sqrt{n}\bigl(\mathbb{E}^{*}[F^{(b)}_n(x)] -
    F(x; \theta_0^*)\bigr)
  \]
  is the bias term, and
  the expected values $\mathbb{E}^{*}[F^{(b)}_n(x)]$ and
$\theta_0^* = \mathbb{E}^{*}[\hat\theta^{(b)}_n]$ can be
approximated by, respectively,
$\mathbb{E}_B^{*}[F^{(b)}_n(x)] = \frac{1}{B}\sum_{b = 1}^B F^{(b)}_n(x)$ and
$\mathbb{E}_B^{*}[\hat\theta^{(b)}_n]  =  \frac{1}{B}\sum_{b = 1}^B\hat\theta^{(b)}_n$.
\end{enumerate}
The p-value of the block bootstrap KS test can be approximated
as $p = \sum_{b=1}^B \mathbf{1}\{T^{(b)}_n > T_n\} / B$.

Assume that the time series $\{X_i\}$ is stationary and
strongly mixing. Using arguments similar to those
in \citet{kunsch1989jackknife}, under certain conditions, the
following results can be shown to
justify the validity of the NPBB procedure of the KS test.
Let $f(x; \theta) = {\partial F(x; \theta)} / {\partial \theta}$
and $f(x, \theta_0)$ be its evaluation at $\theta = \theta_0$.
Let $d$ be the  dimension of~$\theta$.
As $1/l+l/n\rightarrow 0$,
\[
  T_n \overset{d}{\rightarrow}
  \sup_x \big\vert G_{\infty}(x)+f^\top (x;\theta_0)U \big\vert
\]
for some appropriately defined Gaussian process $G_{\infty}(\cdot)$
and $d$-dimensional Gaussian random vector $U$.
Further,  as $1/l+l/n\rightarrow 0$,
\[
  \big\vert \mathbb{P}(T_n^{(b)}\leq x \mid \{X_i: i = 1, \ldots, n\}) -
  \mathbb{P}(T_n\leq x) \big\vert
  \overset{a.s.}{\rightarrow} 0
\]
for any fixed $x$, where $T_n^{(b)}$ is the statistic computed
based on the bootstrap sample $\{X_i^{(b)}: i = 1, \ldots, n\}$.
Following the arguments in the proof of the Glivenko--Cantelli
Theorem, the bootstrap consistency result could be further
strengthened to allow uniform convergence over $x$. That is,
as $1/l+l/n\rightarrow 0$,
\[
  \sup_x \big \vert
  \mathbb{P}(T_n^{(b)}\leq x \mid \{X_i: i = 1, \ldots, n\}) - \mathbb{P}(T_n\leq x) \big\vert
  \overset{a.s.}{\rightarrow} 0.
\]

\subsection{Heuristic justification for the NPBB bias correction}

A clear understanding of the bias correction term $K_n(x)$ is 
essential for appreciating how the NPBB procedure effectively replicates 
the null distribution of the KS statistic in the presence of serial 
dependence and estimated parameters. When the parameters~$\theta$ of a
hypothesized distribution $F(x;\theta)$ are estimated from
the data with $\hat\theta_n$, the null distribution of the KS statistic 
$T_n = \sup_x |\sqrt{n}(F_n(x) - F(x;\hat\theta_n))|$ is altered from
the Kolmogorov distribution under independence and known parameters.
The NPBB procedure aims to replicate this complex null distribution.

In the ``bootstrap world,'' each resample yields an empirical distribution 
$F_n^{(b)}$ and a parameter estimate $\hat\theta^{(b)}_n$. The term 
$\sqrt{n}(F_n^{(b)}(x) - F(x;\hat\theta^{(b)}_n))$ is the bootstrap analog of
the goodness-of-fit process. However, this bootstrap process
inherently possesses its own mean or systematic deviation. The bias
correction $K_n(x)$ estimates this mean.
Specifically, $\mathbb{E}^*[F_n^{(b)}(x)]$ is the average empirical
distribution function over all bootstrap samples, and
$\theta^*_0 = \mathbb{E}^*[\hat\theta^{(b)}_n]$ is the average of 
the parameter estimates from these bootstrap samples. Thus, $K_n(x)$
quantifies the expected difference, in the bootstrap world, between
the empirical distribution function and the hypothesized distribution
function (evaluated at the average bootstrap parameter estimate).
Subtracting $K_n(x)$ from
$\sqrt{n}(F_n^{(b)}(x) - F(x;\hat\theta^{(b)}_n))$ centers the 
bootstrap goodness-of-fit process. This ensures that the variability of the resulting 
bootstrap statistics $T_n^{(b)}$ more faithfully reflects the variability of 
the original $T_n$ statistic around its true but complex null
distribution.

In theory, both $K_n(x)$ and $C_n(x)$ serve the same purpose of
centering the bootstrap goodness-of-fit process, and the circular
block bootstrap reproduces the temporal dependence structure of the
data. The two corrections aims the same centering, differing only in
whether the centering uses bootstrap expectations or sample-based
estimates. Additional simulations (see Section~S-1 of the Supplement)
show that, across all dependence levels and sample sizes considered,
the empirical sizes and powers were nearly identical. For the settings
studied, the two corrections performed nearly identically; we retain
$K_n(x)$ for conceptual clarity and consistency of presentation.

\section{Simulation Studies}\label{sec:simu}

In this simulation study, our objective is twofold: firstly, to demonstrate that
under the null distribution, our test rejects the null hypothesis ($H_0$)
approximately at the specified size or significance level. Secondly, we aim to
illustrate that under the alternative hypothesis, our test rejects
$H_0$ with substantial power. The fulfillment of both criteria would indicate
the method's efficacy.

\subsection{Size}
To assess the actual size of the test, our strategy is to
generate a large number of stationary time series $X_i$ with a certain marginal
distribution $F(\cdot; \theta)$, and for each time series, use our method to
test if $X_i$ follows a hypothesized marginal distribution $F(\cdot; \theta)$
with some unknown $\theta$. If the test holds its size, the p-value
of the test should be uniformly distributed between 0 and 1. We will validate
our method with different marginal distributions to ensure that it is robust.
In order for the method to work, a large number of replicates is necessary.

We generated time series with marginal distributions $N(8, 8)$ and
$\Gamma(8, 1)$ with seven levels of Kendall's
$\tau \in \{-.75, -.50, -.25, 0, .25, .50, .75\}$, and
sample size $n \in \{100, 200, 400, 800\}$. Kendall's $\tau$ was chosen as a
measure of serial dependence between $X_i$ and
$X_{i+1}$ as it is invariant to monotone marginal
transformations.
To generate the samples to which our
method would be applied, we simulated a time series $W_i$ from a 1st
order autoregressive or AR(1) process:
\begin{equation*}
W_i = \phi W_{i-1} + \epsilon_i,
\end{equation*}
where $\phi$ is an autoregressive coefficient, and $\epsilon_i$ is a series of
independent errors from a Normal distribution with mean zero and variance
$\sigma_{\epsilon}^2$. The strength of the serial dependence is controlled by
$\phi$, which was set to seven levels:
$\{-0.924, -0.707, -0.383, 0, 0.383, 0.707, 0.924\}$, as these
correspond to the desired values for $\tau$. The
series $W_i$ has mean zero and variance
$\sigma_x^2 = \sigma_{\epsilon}^2 / (1 - \phi^2)$, so for each value of
$\phi$,  we
set $\sigma_{\epsilon}^2 = (1 - \phi^2)$ such that $\sigma_x^2 = 1$.
First, we generated a
marginal $N(8, 8)$ by marginally transforming $W_i$ by
\begin{equation*}
X_i = F^{-1}[\Phi(W_i)],
\end{equation*}
where $F^{-1}(p)$ is the quantile function for the $N(8, 8)$
distribution and $\Phi$ is the distribution function of the standard Normal
distribution.
Then we generated a marginal Gamma series by the same procedure, but
replacing $F^{-1}(p)$ with the quantile function for the $\Gamma(8, 1)$
distribution. After the transformation
to the marginal Gamma distribution, the lag-1 autocorrelations are approximately
\[
  \{-0.876, -0.674, -0.368, 0, 0.377, 0.701, 0.922\}.
\]

These distributions
were chosen to compare results on Normal and non-Normal
error structures. The specific parameters were chosen because the distributions
are very similar and their first two moments are the same. For the purposes of
evaluating if the test holds it size, when $X_i \sim N(8, 8)$, we tested that the
marginal distribution is from the
Normal family, or
$X_i \sim N(\cdot ; \mu, \sigma^2), \quad i = 1, \ldots, n$
for some unspecified $\mu$ and $\sigma$. When $X_i \sim \Gamma(8, 1)$, we tested
that the marginal distribution family is from the Gamma family,
or
$X_i \sim \Gamma(\cdot ; \alpha, \beta), \quad i = 1, \ldots, n$.
for some unspecified $\alpha$ and $\beta$. For the block bootstrap step,
we used $B = 1000$ and $l = \lceil n^{1/3} \rceil$.
As noted in Subsection~\ref{sub:npbb}, previous literature
suggests varying block size based on the dependence structure of the series.

For this reason, we also tried implementing our algorithm using the
plug-in approach for circular bootstrap block size introduced in
\citet{politis2004automatic}. Results and a discussion of this
additional simulation study are provided in Section~S-2
in the Supplement. We also compared empirical performance for the
block bootstrap KS test when using the $K_n(x)$ correction
versus the $C_n(x)$. The $K_n(x)$ correction was used
in the main manuscript whereas results and a discussion for the
$C_n(x)$ correction is included in the supplement.
For each setting for $F$, $\tau$, and $n$, we replicate the method
10000 times to get the distribution of the p-values $p_r$ for
$r \in \{1, \dots, 10000\}$
for the test when applied to samples from the same data generating
process.

\begin{figure}[tbp]
  \includegraphics[width = .9\textwidth]{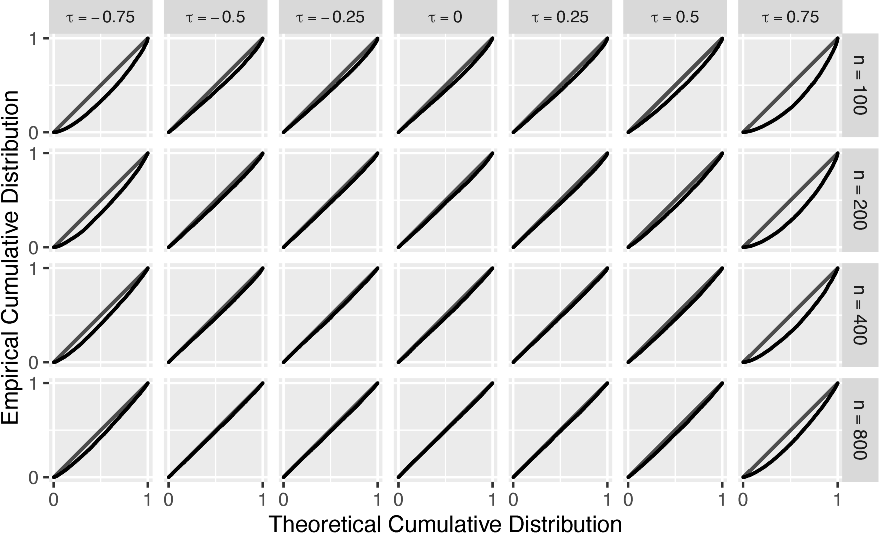}
  \centering
  \caption{Q-Q plots of the p-values testing that a time series
    have marginal Normal distribution with true data generating distribution
    being $N(8,8)$.}
  \label{fig:qq_n}
    \hspace{5cm}
  \includegraphics[width = .9\textwidth]{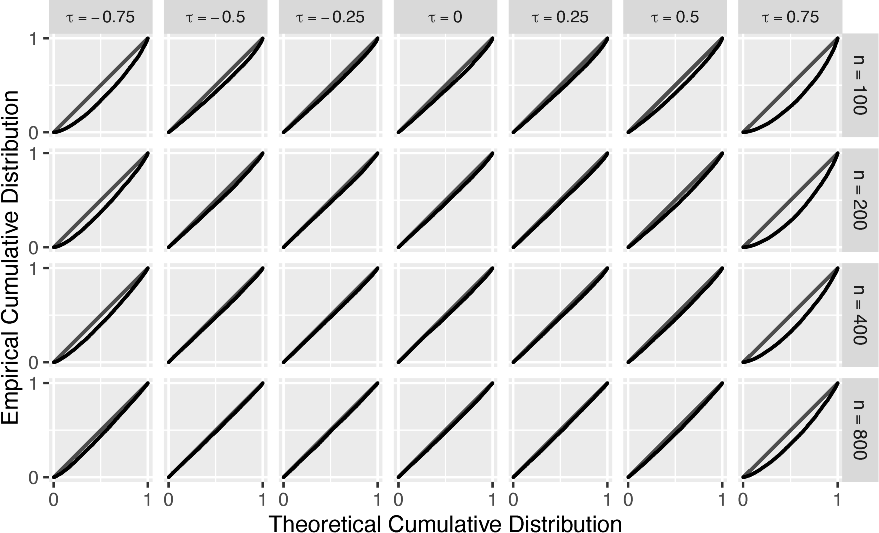}
  \caption{Q-Q plots of the p-values testing that a time series
    have marginal Gamma distribution with true data
    generating distribution $\Gamma(8,1)$.}
  \label{fig:qq_g}
\end{figure}

Figures~\ref{fig:qq_n}--\ref{fig:qq_g} show Quantile-Quantile (Q--Q) plots
of the $p$-values testing the marginal distribution of the stationary time
series generated under the null hypothesis. Figure~\ref{fig:qq_n} corresponds
to the case where the true data generating distribution is $N(8,8)$, and
Figure~\ref{fig:qq_g} corresponds to the case where the true data generating
distribution is $\Gamma(8,1)$. The plots were created using the R packages
\textsl{qqplotr} and \textsl{ggplot2} \citep{qqplotr, ggplot2}. If the
distribution of the p-values is uniform, the points will be aligned with the
diagonal line.

The plots suggest that the extent to which the p-values are uniformly
distributed is dependent on the sample size and the level of serial dependence.
A small sample size like $n = 100$ or $200$ seems adequate when Kendall's
$\tau$ has an absolute value $|\tau| \leq 0.25$. For $|\tau| \geq 0.5$, 
a sample larger than
$n = 200$ seems necessary for the p-values to be uniformly distributed. For
$|\tau| \geq 0.75$, a sample larger than $n = 800$ appears to be
necessary, as the p-values are not aligned with the line in any of the Q-Q
plots. This is not necessarily a cause for concern, as (for Normal margins)
a Kendall's $\tau$ of 0.75 corresponds to a $\phi$ of 0.924, which is very high.
Additionally, positive serial dependence appears to misalign the distribution
of the p-values more so than negative serial dependences of the same strength.
Results are very similar for Normal and Gamma margins.

\begin{table}[tbp]
\centering
\caption{Empirical sizes or rejection rates
for the test that sample follows its true distribution family for
different values of AR(1) coefficient and for different
significance levels.}
\label{table:size}
\begin{tabular}{ccc ccc c ccc}
  \toprule
$F(\cdot; \theta)$ & $n$ & $\alpha$ & \multicolumn{7}{c}{Kendall's $\tau$}\\
\cmidrule(lr){4-10}
   &  &  & $-0.75$ & $ -0.50$ & $-0.25$ & $0.00$ & $0.25$ & $0.50$ & $ 0.75$ \\
  \midrule
{$N(8,8)$} & {100} & 0.01 & 0.0521 & 0.0124 & 0.0112 & 0.0106 & 0.0114 & 0.0158 & 0.0563 \\
& & 0.05 & 0.1470 & 0.0628 & 0.0545 & 0.0548 & 0.0575 & 0.0725 & 0.1705 \\
& & 0.10 & 0.2277 & 0.1186 & 0.1147 & 0.1072 & 0.1154 & 0.1424 & 0.2714 \\
& {200} & 0.01 & 0.0439 & 0.0113 & 0.0090 & 0.0114 & 0.0109 & 0.0138 & 0.0590 \\
& & 0.05 & 0.1278 & 0.0560 & 0.0510 & 0.0532 & 0.0546 & 0.0691 & 0.1726 \\
& & 0.10 & 0.2080 & 0.1123 & 0.1069 & 0.1086 & 0.1051 & 0.1332 & 0.2650 \\
& {400} & 0.01 & 0.0324 & 0.0107 & 0.0099 & 0.0108 & 0.0097 & 0.0126 & 0.0467 \\
& & 0.05 & 0.1016 & 0.0500 & 0.0497 & 0.0560 & 0.0489 & 0.0592 & 0.1418 \\
& & 0.10 & 0.1734 & 0.1023 & 0.1002 & 0.1089 & 0.1017 & 0.1192 & 0.2232 \\
& {800} & 0.01 & 0.0252 & 0.0090 & 0.0098 & 0.0111 & 0.0102 & 0.0135 & 0.0354 \\
& & 0.05 & 0.0883 & 0.0504 & 0.0486 & 0.0478 & 0.0481 & 0.0618 & 0.1219 \\
& & 0.10 & 0.1555 & 0.1026 & 0.0986 & 0.0961 & 0.0997 & 0.1206 & 0.2044 \\ [1ex]
{$\Gamma(8,1)$} & {100} & 0.01 & 0.0488 & 0.0142 & 0.0117 & 0.0124 & 0.0111 & 0.0161 & 0.0576 \\
  & & .05 & 0.1425 & 0.0641 & 0.0611 & 0.0585 & 0.0589 & 0.0732 & 0.1687 \\
  & & 0.10 & 0.2241 & 0.1224 & 0.1147 & 0.1175 & 0.1150 & 0.1375 & 0.2679 \\
  & {200} & 0.01 & 0.0408 & 0.0100 & 0.0089 & 0.0122 & 0.0116 & 0.0143 & 0.0608 \\
  & & 0.05 & 0.1255 & 0.0559 & 0.0507 & 0.0534 & 0.0558 & 0.0645 & 0.1690 \\
  & & 0.10 & 0.1991 & 0.1105 & 0.1022 & 0.1048 & 0.1072 & 0.1233 & 0.2664 \\
  & {400} & 0.01 & 0.0314 & 0.0097 & 0.0114 & 0.0094 & 0.0103 & 0.0131 & 0.0430 \\
  & & 0.05 & 0.1002 & 0.0482 & 0.0521 & 0.0531 & 0.0561 & 0.0618 & 0.1404 \\
  & & 0.10 & 0.1697 & 0.1025 & 0.1066 & 0.1018 & 0.1100 & 0.1177 & 0.2232 \\
  & {800} & 0.01 & 0.0249 & 0.0092 & 0.0113 & 0.0112 & 0.0100 & 0.0113 & 0.0381 \\
  & & 0.05 & 0.0902 & 0.0488 & 0.0532 & 0.0511 & 0.0543 & 0.0598 & 0.1237 \\
  & & 0.10 & 0.1561 & 0.1024 & 0.1072 & 0.1034 & 0.1067 & 0.1177 & 0.1997 \\
   \bottomrule
\end{tabular}
\end{table}

For a closer scrutiny on whether the test holds its size at commonly used
significance levels 0.01, 0.05, and 0.10, Table~\ref{table:size}
summarizes the empirical sizes of the tests based on 10000 replicates.
The agreement between the empirical sizes and the nominal size
improves as sample size improves, as expected. For $\tau = -0.75$
and $\tau = 0.75$, a sample size as large as 800 is still not sufficient for the
agreement to be reached. For weaker dependence levels, our approach is working
for $-0.25 < \tau < 0.25$ for $n = 100$. However, a larger sample size seems
necessary, greater than 400 perhaps, for $\tau$ as strong as $0.5$. To
summarize, for weaker dependence, our method works without need for a
particularly large sample size, whereas for moderate dependence, a larger sample
size is required. When the agreement is not met, the empirical size is
usually larger than the nominal size, which indicates that the test is too liberal.
In other words, the test rejects the null hypothesis more often than it should.
For practical situations with a $\tau$ of magnitude 0.5 or
lower, it is still reasonable to recommend our method with $n \ge 400$.
This aligns with the high sample size requirement obverved for block
bootstrapping confidence intervals~\citep{chandy2024sample}.

\subsection{Power}
The power of the proposed test is investigated with data generated from
distributions that are not in the hypothesized distribution family.
We generated the same scheme of time series as we did in the size investigation,
but tested that the time series was from the other distribution.
Because the support of $N(8, 8)$ is
$(-\infty, \infty)$, but the support of $\Gamma(8, 1)$ is $(0, \infty)$, we used
the \textsl{truncdist} package \citep{truncdist} to truncate the series at
values less than or equal to 0 when testing that the $N(8, 8)$ sample was
from $\Gamma(8, 1)$. For each setting for $F$, $\tau$, and $n$, we replicate the
method $10000$ times
to obtain a p-value $p_r$ for each $r \in \{1, \ldots, 10000\}$.
The empirical rejection rates were obtained at a nominal significance level 0.05.

\begin{figure}[tbp]
  \centering
  \includegraphics[scale=1]{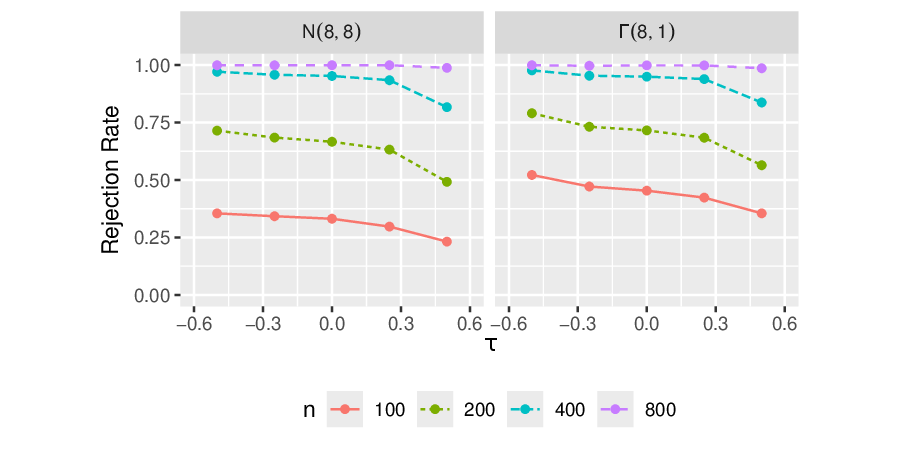}
  \caption{Empirical power curve as a function of $\tau$ for
    $n \in \{100, 200, 400, 800\}$ and true marginal distribution
    $\in \{N(8,8), \Gamma(8,1)\}$. When the data was generated from $N(8,8)$,
    we tested for the Gamma family. When the data was generated from
    $\Gamma(8,1)$, we tested for the Normal family.
  }
  \label{fig:rr}
\end{figure}

Figure~\ref{fig:rr} displays the empirical power curve of the test at
significance level 0.05 as the lag-1 Kendall's~$\tau$ increases from
$-0.5$ to $0.5$. We omitted $-0.75$ and 0.75 from this study because we already
found that our method doesn't work under the null hypothesis for such high
dependence levels under the sample sizes considered. It appears that the
rejection rate seems to decline as $\tau$
increases. Rejection rates are negatively affected by positive serial
dependence, but they do not seem to be affected by negative serial dependencies
of the same strength.
Using $n \geq 400$ seems to be good enough
if $\tau < 0.5$, but using a $n \geq 800$ will result in
rejection rates close to 1 for $-0.5 < \tau < 0.5$. 
Results for $n \geq 400$
do not appear to be to different when comparing Gamma and Normal
margins, indicating that the method is robust to the marginal distribution.
The rejection rates for $n = 800$ are very
high (almost 1), indicating that under $H_A$, the test is very powerful
when a large enough sample size is used.

\section{A Real Data Example}
\label{sec:real}

Using the \textsl{tseries} package \citep{tseries},
we gathered daily closing prices for the S\&P 500 index from January 1st, 2020
to December 31st, 2023. Daily returns were obtained by taking the difference in
the logarithm of the closing prices.
It is known as stylized facts that daily return series often exhibit nonlinear
dependence and unconditional heavy tails \citep[e.g.,][]{ryden1998stylized,
  cont2001empirical}. The lag-1 Kendall's $\tau$ of the series is
about $-0.0298$, indicating a fairly weak serial linear correlation, but
certainly not meaning absence of serial dependence.
Using our method, we tested that the daily returns are normally
distributed and Student's $t$ distributed
with degrees of freedom $\nu \in \{30, 20, 10, 5, 4, 3, 2, 1\}$
with the proposed NPBB method.
For comparison, we tested the same hypotheses using 1) the
SPB method of \citet{zeimbekakis2024misuses},
which accounts for both serial dependence and unspecified parameters;
2) NPB bias correction method of
\citet{babu2004goodness}; and 3) the parametric bootstrap (PB) method shown
by \citet{zeimbekakis2024misuses}. The latter of the two methods discard the
serial dependence. In all methods, we used $B = 10,000$ bootstrap replicates.
The p-values for these tests are summarized in
Table~\ref{table:SP5004}.

\begin{table}[ht]
\centering
\caption{P-values for testing marginal Student's $t$ distribution on
  4 years of daily returns of S\&P 500 with $B = 10,000$
  using different degrees of freedom~$\nu$.
  NPBB: nonparametric block bootstrap;
  SPB: semiparametric bootstrap;
  NPB: nonparametric bootstrap;
  PB: parametric bootstrap.}
\label{table:SP5004}
\begin{tabular}{ccccc}
  \toprule
$\nu$ & NPBB & SPB & NPB & PB\\
  \midrule
$\infty$ & 0.0001 & 0.0000 & 0.0000 & 0.0000 \\
  30 & 0.0019 & 0.0000 & 0.0000 & 0.0000 \\
  20 & 0.0027 & 0.0000 & 0.0000 & 0.0000 \\
  10 & 0.0125 & 0.0002 & 0.0003 & 0.0003 \\
  5 & 0.0610 & 0.0192 & 0.0222 & 0.0141 \\
  4 & 0.1037 & 0.0733 & 0.0574 & 0.0499 \\
  3 & 0.3133 & 0.3000 & 0.2759 & 0.2686 \\
  2 & 0.0418 & 0.0461 & 0.0377 & 0.0443 \\
  1 & 0.0000 & 0.0000 & 0.0000 & 0.0000 \\
  \bottomrule
\end{tabular}
\end{table}

While differences in p-values among different methods are expected, our analysis
reveals some instances where conclusions at the 0.05 significance level
using our method differ from those obtained using the methods proposed by
\citet{babu2004goodness} and \citet{zeimbekakis2024misuses}.
For example, our NPBB method fails to reject the
hypothesis that the series follows a Student's $t$ distribution with degrees of
freedom $\nu = 5$ with a p-value of 0.061. In contrast,
the SPB method yields a p-value of 0.0192,
the NPB method yields a p-value of 0.0222, and the PB yields a p-value of
0.0141. Although such differences are not guaranteed, this
instance highlights a disagreement in conclusions among methods for the S\&P 500
index. Because NPB and PB both discard serial dependence, it is expected that
their conclusions could be different from our method. It is worth noting that
the SPB result is also not reliable here because the working dependence model is
in the class of ARMA dependence as specified in \citet{zeimbekakis2024misuses},
which does not cover nonlinear dependence usually captured by autoregressive
conditional heteroskedasticity \citep{engle1995arch}. The NPBB method has
no need to specify the serial dependence and, hence, produces results under
weakest assumptions.

\section{Discussion}\label{sec:conclusion}

The KS test remains a crucial tool for researchers across diverse fields,
enabling critical assessments of population distributions. We introduced a
NPBB variant of the KS test tailored for data with
serial dependencies, which eliminates the need for explicit modeling of serial
dependence in addition to parameter specification within the hypothesized
distribution. Our simulations have demonstrated that, given
sufficiently large samples of time series data and moderate serial dependencies,
the test maintains its size under the null hypothesis and have considerable
power in detecting deviations from the hypothesized marginal distribution. The
method's performance improves as the strength of temporal dependence decreases
or the sample size increases. For moderate serial dependence levels in practice,
the test can be recommended for sample size~$n = 400$ or higher.
Overall, the test provides an analog of the bias-corrected NPB KS test of
\citet{babu2004goodness} in the context of stationary time
series with unspecified serial dependence. Compared to the SPB method
of \citet{zeimbekakis2024misuses}, which requires a working dependence model, it
is reliable with no need to worry about whether the working dependence
model approximates the true dependence model. We note that while the
expectation centering in our paper removes a minor bias term, it is identical
in asymptotic performance to \citet{babu2004goodness}'s correction under 
short-range dependence.

The choice of block size strongly influences the performance of
block-bootstrap methods for dependent data.
As pointed out by \citet{hall1995blocking} and
\citet{lahiri1999theoretical}, the optimal block size depends on the
strength of serial dependence, and using a uniform choice across
different dependence settings can lead to size distortion. Several
data-driven procedures for selecting block size have been proposed,
including plug-in and cross-validation methods
\citep{buhlmann2002bootstraps, politis2004automatic}. A more
comprehensive treatment is given in
\citet{lahiri2013resampling}. Following these suggestions, we applied
the automatic block size selection method of
\citet{politis2004automatic} for circular block bootstrap in our
context. As reported in the supplement, the method frequently selected
block sizes that were overly large—sometimes exceeding the sample
size—especially under strong serial dependence. This reduced the
variability of the bootstrap replicates and affected the reliability
of the test. These observations highlight that while automatic methods
are attractive in principle, their performance in small to moderate
samples and in goodness-of-fit settings like ours may be limited.

Several promising directions for further research emerge from
this study. Firstly, while we used circular block bootstrap, many
other bootstrap variants for serially dependent data, such as the
stationary bootstrap \citep{politis1994stationary} and the dependent wild
bootstrap \citep{shao2010dependent}, could be evaluated under our
correction framework. Secondly, analogous methodologies could be
developed for other goodness-of-fit tests based on empirical
distributions, including the Anderson--Darling and Cramér--von
Mises tests, which are known to outperform the KS test in certain scenarios
\citep{stephens2017tests}. Applying block bootstrap bias correction to
these tests may be particularly beneficial in contexts like extreme
value modeling, where serial dependence is common. Third, it would be
intriguing to extend the proposed NPBB
one-sample KS test to a two-sample setting, allowing for comparison of
marginal distributions without specifying serial dependencies in
either sample. Finally, a rigorous theoretical justification of the
NPBB procedure, including precise conditions for the asymptotic
behavior of the test statistic and the consistency of the bootstrap
approximation, would offer valuable practical insights into the
scenarios where the NPBB approach is expected to perform optimally.
In particular, while the
correction term is designed to calibrate the null distribution of the
test statistic, studying its behavior under alternatives, including
contiguous alternatives, would further inform the power properties of
the method and remains an interesting direction for future research.

\bigskip
\begin{center}
{\large\bf SUPPLEMENTARY MATERIAL}
\end{center}

\begin{description}

\item[Funding statement:]JY and XZ's research was partially supported by NSF grant
DMS2210735.

\item[Conflict of interest disclosure:]The authors declare no potential conflict
of interests.

\item[Data availability statement:] The data and code used in the real data analysis and simulation
studies are publicly available in a GitHub repo:\\
\url{https://github.com/mathewchandy03/Bootstrap_KS_Test}.

\item[Additional simulation results:] Simulation studies on comparison
of two bias corrections and block size selection are reported in an
online pdf Supplement.

\end{description}

\bibliographystyle{asa}
\bibliography{citations}
\end{document}